\documentclass{elsart}
\usepackage{epsfig}

\begin{document}

\begin{frontmatter}

\title{\Large\bf Measurement of Branching Ratios for
{\boldmath $\eta_c$} Hadronic Decays}

\date{29 Aug 2003}

\maketitle

\begin{center}

J.~Z.~Bai$^1$,        Y.~Ban$^{10}$,         J.~G.~Bian$^1$,
X.~Cai$^{1}$,         J.~F.~Chang$^1$,
H.~F.~Chen$^{16}$,    H.~S.~Chen$^1$,   
H.~X.~Chen$^{3}$,
Jie~Chen$^{9}$,        J.~C.~Chen$^1$
% ...
\end{center}
\vskip 0.3cm
\begin{center}
(The BES Collaboration)
\end{center}
\vskip 0.3cm

\small
\begin{center}
$^1$ Institute of High Energy Physics, Beijing 100039, People's Republic of
     China\\
$^2$ China Center of Advanced Science and Technology, Beijing 100080,
     People's Republic of China\\

$^3$ Guangxi Normal University, Guilin 541004, People's Republic of China\\

$^4$ Henan Normal University, Xinxiang 453002, People's Republic of China\\
$^5$ Huazhong Normal University, Wuhan 430079, People's Republic of China\\
$^6$ Hunan University, Changsha 410082, People's Republic of China\\
                                                    
$^7$ Liaoning University, Shenyang 110036, People's Republic of China\\

$^{8}$ Nanjing Normal University, Nanjing 210097, People's Republic of China\\

$^{9}$ Nankai University, Tianjin 300071, People's Republic of China\\
$^{10}$ Peking University, Beijing 100871, People's Republic of China\\
$^{11}$ Shandong University, Jinan 250100, People's Republic of China\\
$^{12}$ Shanghai Jiaotong University, Shanghai 200030, 
        People's Republic of China\\
$^{13}$ Sichuan University, Chengdu 610064,
        People's Republic of China\\                                    
$^{14}$ Tsinghua University, Beijing 100084, 
        People's Republic of China\\
$^{15}$ University of Hawaii, Honolulu, Hawaii 96822\\
$^{16}$ University of Science and Technology of China, Hefei 230026,
        People's Republic of China\\
$^{17}$ Wuhan University, Wuhan 430072, People's Republic of China\\
$^{18}$ Zhejiang University, Hangzhou 310028, People's Republic of China\\
$^{\dagger}$ Visiting professor to University of Michigan, Ann Arbor, MI
48109, USA

\vspace{0.2cm}

\end{center}

\normalsize

\begin{abstract}
In a sample of 58 million $J/\psi$ events collected with the
BES II detector, the process J/$\psi\to\gamma\eta_c$ is
observed in five decay channels: $\eta_c \to
K^+K^-\pi^+\pi^-$, $\pi^+\pi^-\pi^+\pi^-$,
$K^\pm K^0_S \pi^\mp$ (with $K^0_S\to\pi^+\pi^-$),
$\phi\phi$ (with $\phi\to K^+K^-$) and $p\bar{p}$.  From these signals,
we determine \\
$Br(J/\psi\to\gamma\eta_c)\times Br(\eta_c\to K^+K^-\pi^+\pi^-)$       $=(1.5\pm0.2\pm0.2)\times10^{-4}$, \\
$Br(J/\psi\to\gamma\eta_c)\times Br(\eta_c\to \pi^+\pi^-\pi^+\pi^-)$   $=(1.3\pm0.2\pm0.4)\times10^{-4}$, \\
$Br(J/\psi\to\gamma\eta_c)\times Br(\eta_c\to K^\pm K_{S}^{0}\pi^\mp)$ $=(2.2\pm0.3\pm0.5)\times10^{-4}$, \\
$Br(J/\psi\to\gamma\eta_c)\times Br(\eta_c\to \phi\phi)$               $=(3.3\pm0.6\pm0.6)\times10^{-5}$ and \\
$Br(J/\psi\to\gamma\eta_c)\times Br(\eta_c\to p\bar{p})$               $=(1.9\pm0.3\pm0.3)\times10^{-5}$. \\

\vspace{3\parskip}
\noindent{\it PACS:} 13.25.Gv, 14.40.Gx, 13.40.Hq

\end{abstract}

\end{frontmatter}
\clearpage

Hadronic decays of the $\eta_c$ have been studied by Mark
III~\cite{mark3,mark3-2}, DM2~\cite{dm2}, and other
experiments [4-7]. However, the branching fractions of the $\eta_c$ still
have very large errors in the Particle Data Group (PDG)
compilation~\cite{pdg2002}. More recently the branching fractions for
$B\to\eta_c K$ decays and $B\to\eta_c K^*$ have been measured by the
Belle~\cite{belle1,belle2} experiment, and their measured branching
fraction for $\eta_c\to\phi\phi$ is smaller than the PDG
value~\cite{pdg2002}.

In a previous paper~\cite{besetac}, based on 58 million $J/\psi$
events collected in the Beijing Spectrometer (BES II) detector at the
Beijing Electron-Positron Collider, we measured the $\eta_c$ mass and
width using the processes J/$\psi\to\gamma\eta_c$, $\eta_c\to$
$K^+K^-\pi^+\pi^-$, $\pi^+\pi^-\pi^+\pi^-$, $K^\pm K^0_S \pi^\mp$
(with $K^0_S\to\pi^+\pi^-$), $\phi\phi$ (with $\phi\to K^+K^-$) and
$p\bar{p}$, and obtained
$m_{\eta_c}=2977.5\pm1.0 \;(\mbox{sta})\pm1.2 \;(\mbox{sys})$ MeV and
$\Gamma_{\eta_c} = 17.0\pm3.7 \;(\mbox{sta})\pm7.4 \;(\mbox{sys})$ MeV.  In
this paper, we report measurements of the branching ratios for the same
processes.

BES is a conventional solenoidal magnet detector that
is described in detail in Ref.~\cite{bes2}; BESII is the upgraded version
of the BES detector~\cite{bes}. A 12-layer vertex chamber (VTC)
surrounding the beam pipe provides trigger information. A
forty-layer main drift chamber (MDC), located radially outside the
VTC, provides trajectory and energy loss ($dE/dx$) information for
charged tracks over $85\%$ of the total solid angle with a
momentum resolution of $\sigma _p/p = 0.0178 \sqrt{1+p^2}$ ($p$ in
$\hbox{\rm GeV}/c$) and a $dE/dx$ resolution for hadron tracks of
$\sim 8\%$. An array of 48 scintillation counters surrounding the
MDC measures the time-of-flight (TOF) of charged tracks with a
resolution of $\sim 200$ ps for hadrons.  Radially outside the TOF
system is a 12 radiation length, lead-gas barrel shower counter
(BSC).  This measures the energies of electrons and photons over
$\sim 80\%$ of the total solid angle with an energy resolution of
$\sigma_E/E=21\%/\sqrt{E}$ ($E$ in GeV).  Outside the solenoidal
coil, which provides a 0.4~Tesla magnetic field over the tracking
volume, is an iron flux return that is instrumented with three
double layers of counters that identify muons of momentum greater
than 0.5~GeV/c.

A Geant3 based Monte Carlo, SIMBES, which simulates the
detector response, including interactions of secondary particles
in the detector material, is used in this analysis.
Reasonable agreement between data and Monte Carlo simulation is
observed in various channels tested, including $e^+e^-\to(\gamma)
e^+ e^-$, $e^+e^-\to(\gamma)\mu\mu$, $J/\psi\to p\bar{p}$,
$J/\psi\to\rho\pi$ and $\psi(2S)\to\pi+\pi^- J/\psi$,
$J/\psi\to l^+ l^-$.

The event selection criteria for each channel are described in detail
in our previous paper \cite{besetac}. Here we repeat
only the essential information and emphasize those considerations
that are unique to the $\eta_c$ branching ratio measurement.

Candidate events are required to have the correct number of
charged tracks for a given hypothesis. Events are kinematically
fitted with four constraints (4C) to the hypotheses: $J/\psi \to
\gamma K^+K^-\pi^+\pi^-$, $J/\psi \to \gamma
\pi^+\pi^-\pi^+\pi^-$, $J/\psi \to \gamma K^\pm\pi^\mp\pi^+\pi^-$,
and $J/\psi \to \gamma p \bar{p}$. A one-constraint (1C) fit is
performed for the $J/\psi \to \gamma_{miss} K^+K^-K^+K^-$
hypothesis, where $\gamma_{miss}$ indicates that this photon is
not detected. Events with a $\chi^2$ less than 40.0 for a
particular channel are selected.

In order to remove backgrounds from non-radiative decay channels,
all selected events are subjected to (4C)  kinematic fits
to the hypotheses:
$J/\psi \to K^+K^-\pi^+\pi^-$,
$J/\psi \to \pi^+\pi^-\pi^+\pi^-$, and
$J/\psi \to K^\pm\pi^\mp\pi^+\pi^-$ and are
required to satisfy
$\chi^2(J/\psi \to K^+K^-\pi^+\pi^-)>20.0$ (for $K^+K^-\pi^+\pi^-$);
$\chi^2(J/\psi \to \pi^+\pi^-\pi^+\pi^-)>10.0$ (for
$\pi^+\pi^-\pi^+\pi^-$) and
$\chi^2(J/\psi \to K^\pm\pi^\mp\pi^+\pi^-)>10.0$ (for $K^\pm
K_{S}^{0}\pi^\mp$).
For the $J/\psi \to \gamma p \bar{p}$ channel,
we require that the opening angle of the two charged tracks is smaller than
$179^\circ$.
A detailed Monte Carlo simulation shows that these cuts, referred to below as the
$J/\psi$ veto, do not distort
the invariant mass distributions around the $\eta_c$ signal peak.

\begin{figure}[htbp]
  \centering
  \begin{minipage}[b]{0.48\linewidth}
    \centering
    \includegraphics[width=6.5cm]{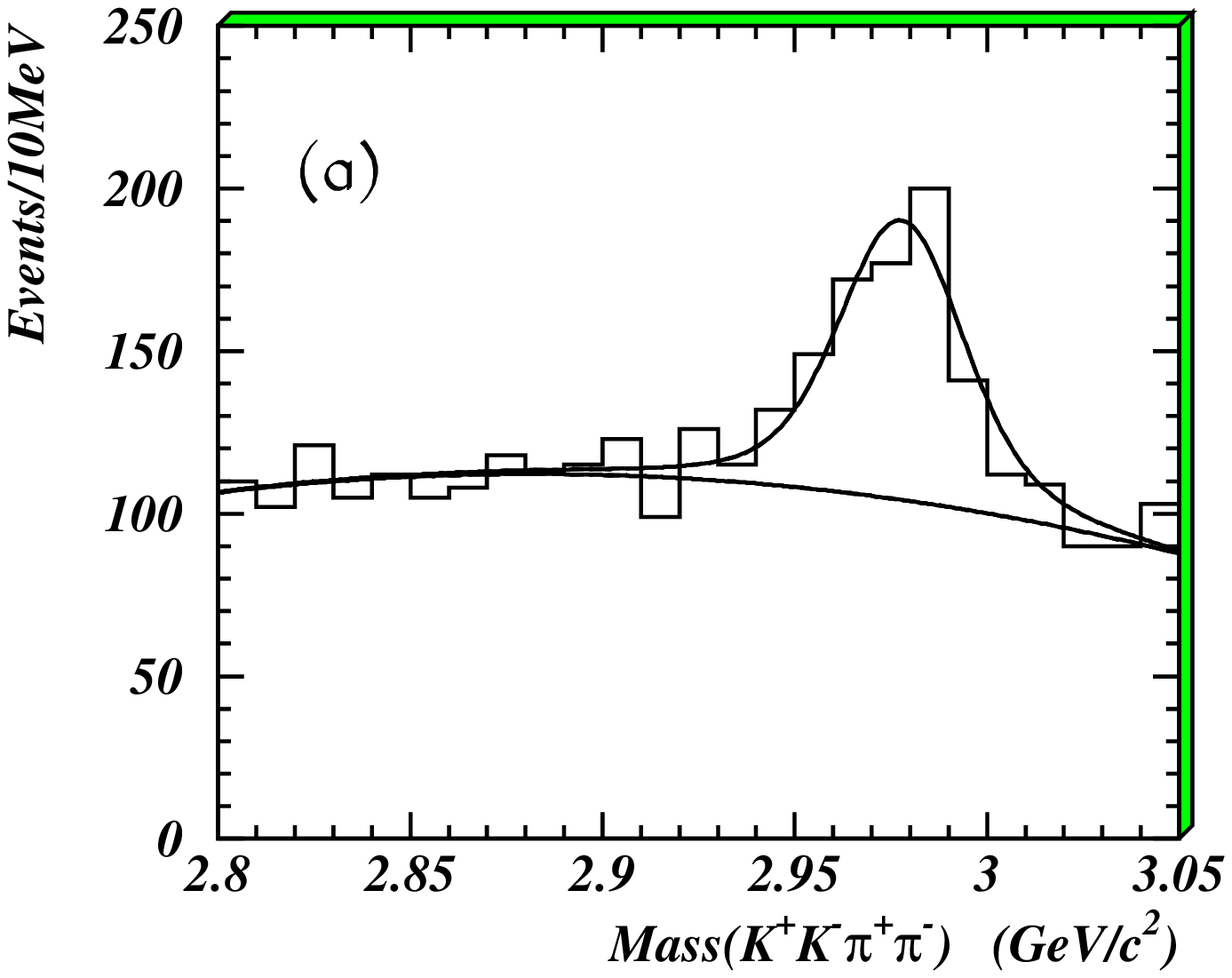}
  \end{minipage}
  \hfill
  \begin{minipage}[b]{0.48\linewidth}
    \centering
    \includegraphics[width=6.5cm]{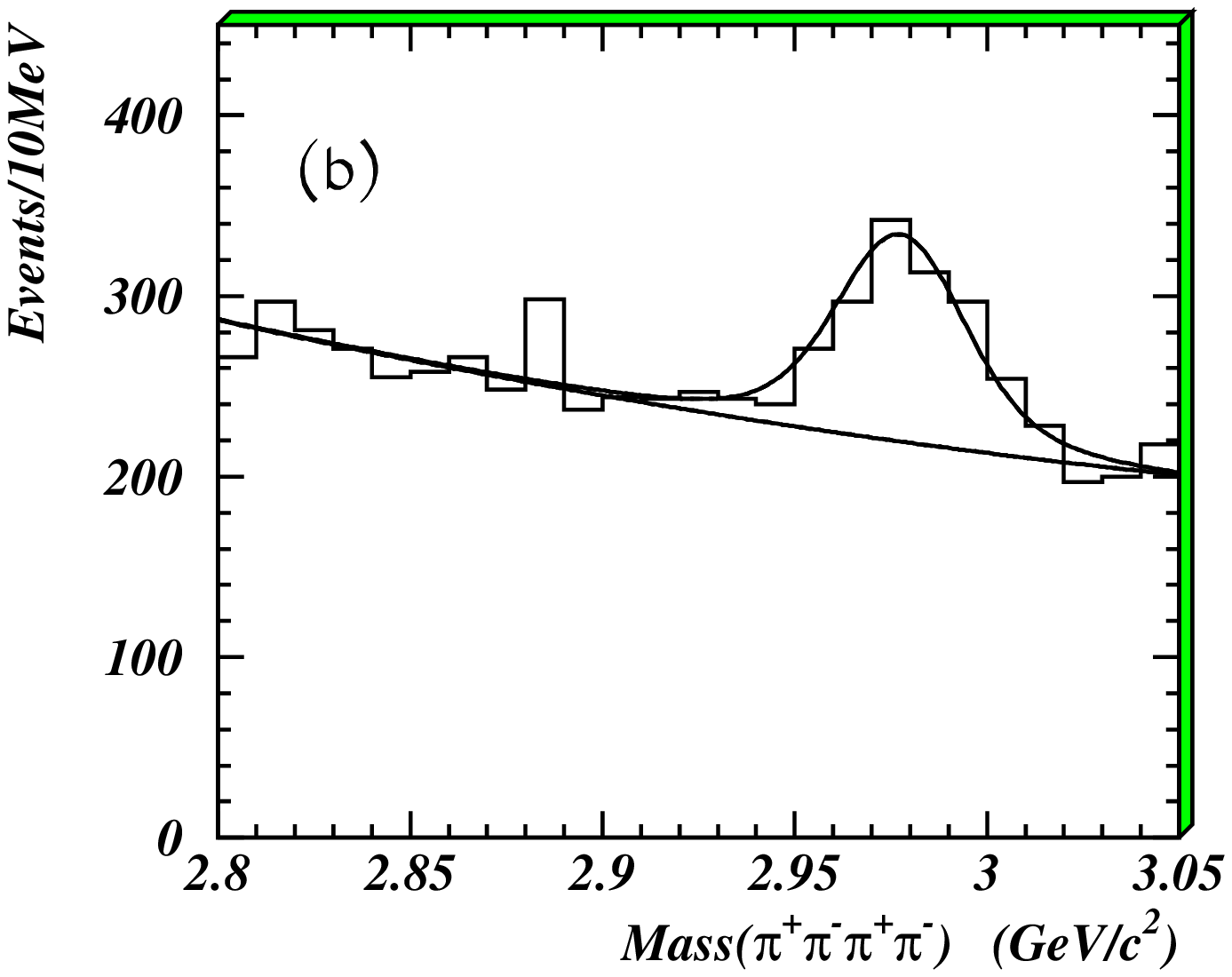}
  \end{minipage}

  \begin{minipage}[b]{0.48\linewidth}
    \centering
    \includegraphics[width=6.5cm]{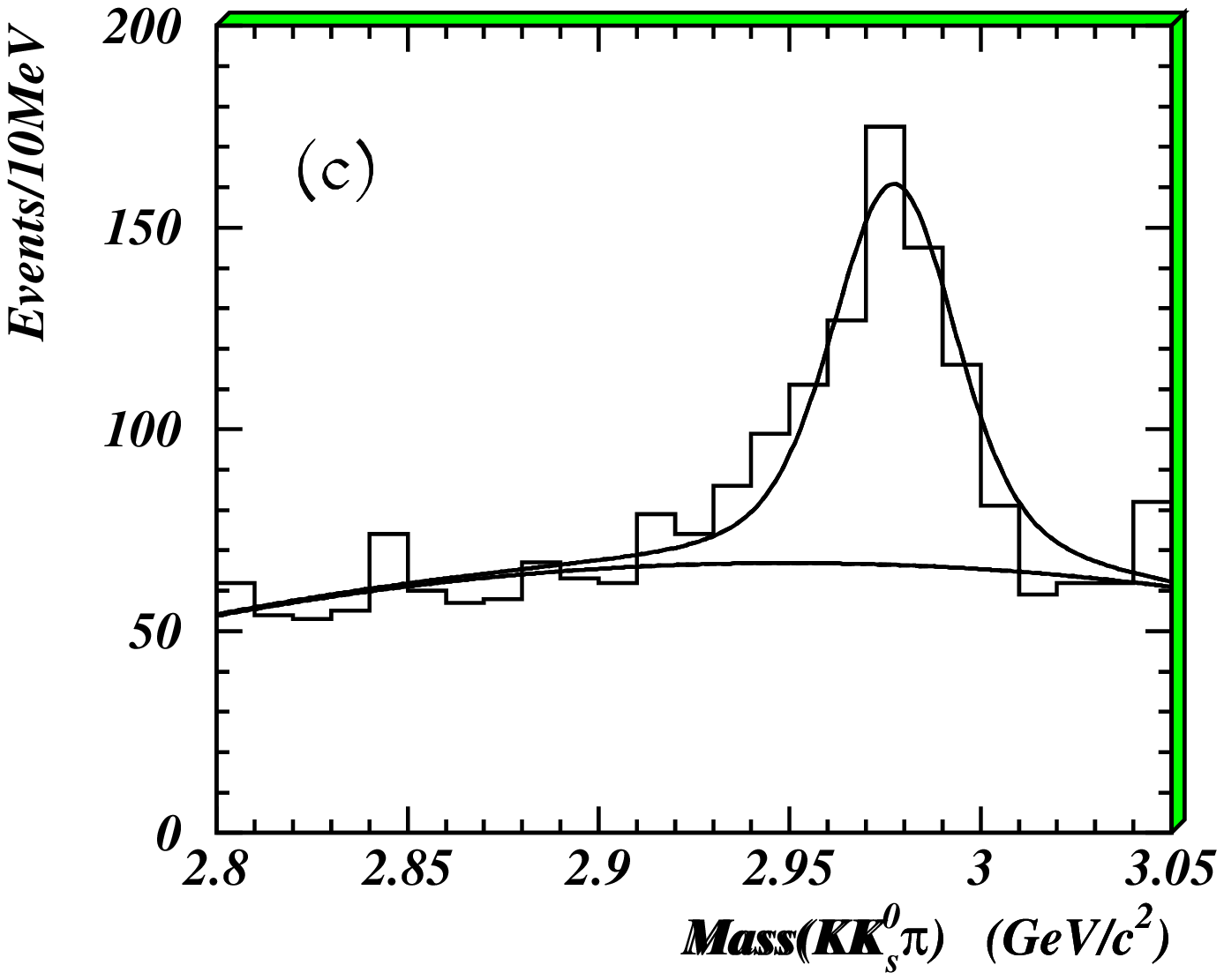}
  \end{minipage}
  \hfill
  \begin{minipage}[b]{0.48\linewidth}
    \centering
    \includegraphics[width=6.5cm]{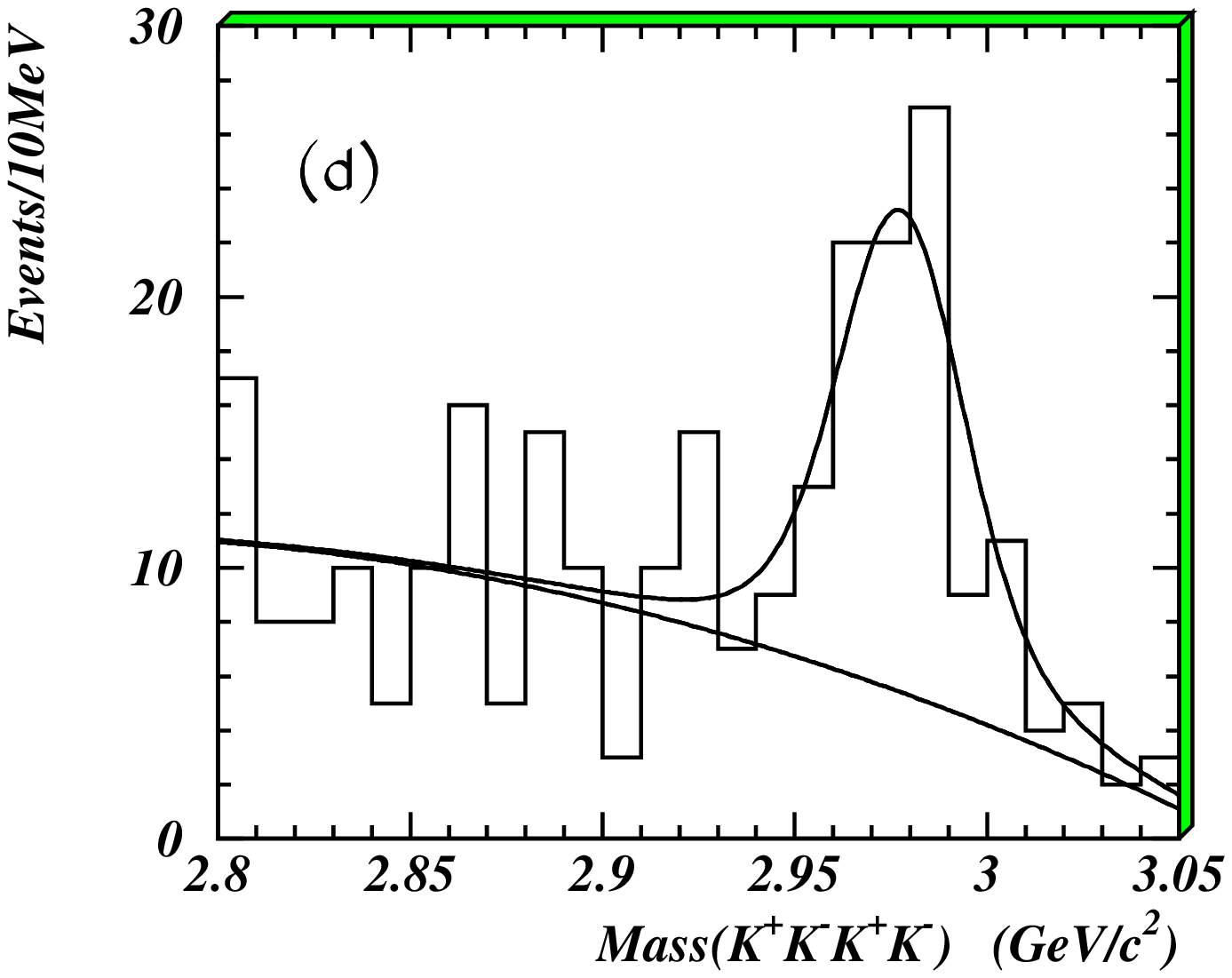}
  \end{minipage}
  \begin{minipage}[b]{0.48\linewidth}
    \centering
    \includegraphics[width=6.5cm]{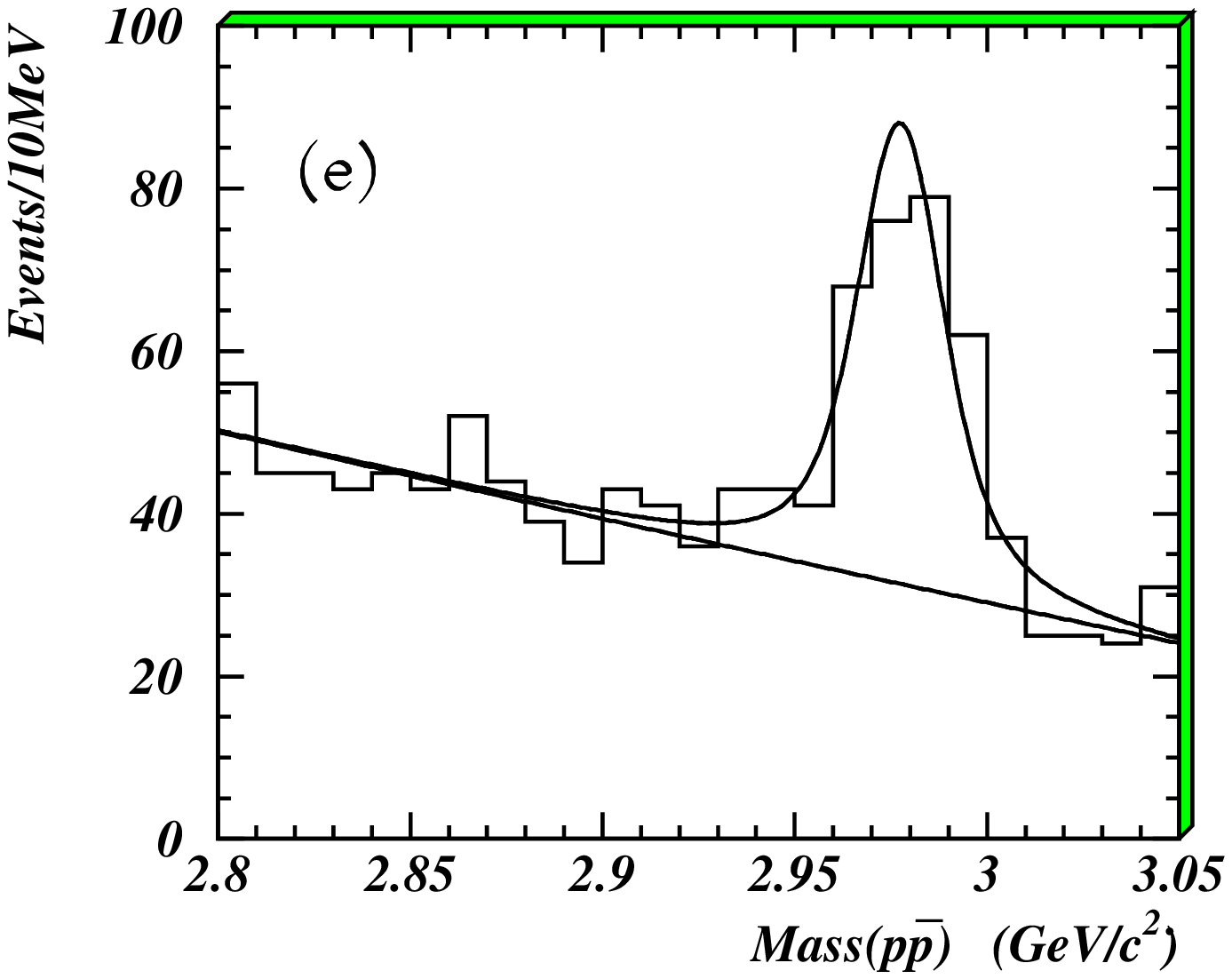}
  \end{minipage}
    \caption[]{Invariant mass distributions in the $\eta_c$ region
               (a) $m_{K^+K^-\pi^+\pi^-}$, (b) $m_{\pi^+\pi^-\pi^+\pi^-}$,
                 (c) $m_{K^\pm K^0_S\pi^\mp}$, (d) $m_{\phi\phi}$ and
                 (e) $m_{p\bar{p}}$. The histograms correspond to the
               data; the curves are the fit result.
}
    \label{fit-2xx-final}
\end{figure}

After event selection, the invariant mass spectra for the
individual decay modes are obtained, as shown in
Fig.~\ref{fit-2xx-final}. An unbinned maximum likelihood fit using
MINUIT~\cite{minuit} is performed to all five channels
simultaneously. The fitting method is described in detail in
our previous paper~\cite{besetac}.

The branching ratio can be calculated using
$$
Br = \frac{N_{fit}/\epsilon}{N_{J/\psi}}  = \frac{N}{N_{J/\psi}},
$$ where $\epsilon$ is the detection efficiency; $N=N_{fit}/\epsilon$
is the efficiency-corrected number of $\eta_c$ events obtained
directly from the fit and corrected using $Br(K^0_s \to \pi^+ \pi^-)$
and $Br(\phi \to K^+ K^-)$~\cite{pdg2002} where necessary; and
$N_{J/\psi}=(57.7\pm2.72)\times 10^6$~\cite{jpsi-n} is the total
number of $J/\psi$ events.  The numbers of $\eta_c$ events determined
from the fit and the corresponding branching ratios, by decay channel,
are listed in Table~\ref{Fit-events}.

\begin{table}[htb]
\caption{Number of $\eta_c$ events and corresponding branching ratios for the individual channels
(corrected using $Br(K^0_s \to \pi^+ \pi^-)$ and $Br(\phi \to K^+ K^-)$~\cite{pdg2002}  where necessary).} \label{Fit-events}
\begin{center}
\begin{tabular}{|l|c|c|c|}
\hline
Process & No. of events & No. of events          & Product of  \\
$J/\psi\to\gamma\eta_c,$    & (detected)      & (efficiency-corrected) &        branching ratios                     \\ \hline
$\eta_c \to K^+K^-\pi^+\pi^-$      & $ 413\pm 54$ & $8453\pm 1110$ &
$(1.5\pm0.2\pm 0.2)\times10^{-4}$ \\ \hline
$\eta_c \to \pi^+\pi^-\pi^+\pi^-$  & $ 542\pm 75$ & $7643\pm 1062$ & $(1.3\pm0.2\pm0.4)\times10^{-4}$ \\ \hline
$\eta_c \to K^\pm K_{S}^{0}\pi^\mp$& $ 609\pm 71$ & $12516\pm1460$ & $(2.2\pm0.3\pm0.5)\times10^{-4}$ \\ \hline
$\eta_c \to \phi\phi$              & $ 357\pm 64$ & $ 1922\pm 357$ & $(3.3\pm0.6\pm0.6)\times10^{-5}$ \\ \hline
$\eta_c \to p\bar{p}$              & $ 213\pm 33$ & $ 1105\pm 171$ & $(1.9\pm0.3\pm0.3)\times10^{-5}$ \\ \hline
\end{tabular}
\end{center}
\end{table}

The main systematic error contributions in measuring the $\eta_c$
branching ratios originate from uncertainties in the background shape
parameterization used, differences between different Monte Carlo
simulations of the drift chamber wire resolution, detection efficiency
differences due to uncertainties in $\eta_c$ decay sequences into the
final state (for $\eta_c \to \pi^+\pi^-\pi^+\pi^-$, $\eta_c \to
K^+K^-\pi^+\pi^-$ and $\eta_c\to K^\pm K_{S}^{0}\pi^\mp$), differences
in the photon efficiency determined using data and that determined
from the Monte Carlo simulation, particle identification
uncertainties, and the uncertainty in the total number of $J/\psi$
events.  In Fig.~\ref{fit-2xx-final}, second-order polynomials are used
to describe the backgrounds.  The systematic errors due the background
shape are studied by using instead linear polynomial functions to fit
the backgrounds in Fig.~\ref{fit-2xx-final}(b), (d), and (e) and third
order polynomials to fit the backgrounds in Fig.~\ref{fit-2xx-final}(a)
and (c), changing the upper fitting bound from 3.05 to 3.07 GeV/$c^2$,
and removing the $J/\psi$ veto from the event selection. The relative
systematic errors from these sources are listed in
Table~\ref{etac-bg}. Since the errors are correlated, we choose the
largest one as the systematic error due to the background shape.
\begin{table}[htbp]
\caption{Relative systematic error caused by background shape.} \label{etac-bg}
\begin{center}
\begin{tabular}{|l|c|c|c|c|c|}
\hline
Sources      & $K^+K^-\pi^+\pi^-$ & $\pi^+\pi^-\pi^+\pi^-$ & $K^\pm K_{S}^{0}\pi^\mp$ & $\phi\phi$ & $p\bar{p}$ \\ \hline \hline
background polynomial &  4.4\%  & 7.6\%    & 2.5\%    & 8.3\%    & 3.2\%  \\
fitting range       &  9.4\%  & 8.4\%    & 17.2\%   & 15.5\%   & 10.6\% \\
$J/\psi$ veto       &  1.7\%  &26.6\%    & 10.1\%   & 17.3\%   & 15.2\% \\
\hline
\end{tabular}
\end{center}
\end{table}

The relative systematic errors for the individual channels
are summarized in Table~\ref{etac-sys}, where the individual
contributions are added in quadrature to obtain the total relative
systematic error.  The systematic errors on the product branching
ratios are given in Table~\ref{Fit-events}.

\begin{table}[htbp]
\caption{Relative systematic error summary.} \label{etac-sys}
\begin{center}
\begin{tabular}{|l|c|c|c|c|c|}
\hline Sources      & $K^+K^-\pi^+\pi^-$ & $\pi^+\pi^-\pi^+\pi^-$ & $K^\pm K_{S}^{0}\pi^\mp$ & $\phi\phi$ & $p\bar{p}$ \\ \hline \hline
BG shape            &  9.4\%  & 26.6\%    & 17.2\%    & 17.3\%    & 15.2\% \\ \hline
wire resolution     & 10.4\%  & 17.1\%    & 13.1\%    &  2.9\%    &  4.7\% \\ \hline
$\eta_c$ decay sequences & 4.5\%  & 4.5\% & 1.0\%     &  -    &  - \\ \hline
$\gamma$ efficiency &  2.0\%  &  2.0\%    &  2.0\%    &  2.0\%    &  2.0\% \\ \hline
particle ID         &  2.5\%  &  2.7\%    &  2.2\%    &  2.5\%    &  1.1\% \\ \hline
N$_{J/\psi}$        &  4.7\%  &  4.7\%    &  4.7\%    &  4.7\%    &  4.7\% \\ \hline\hline
Total               & 15.8\%  & 32.5\%    & 22.3\%    & 18.4\%    & 16.7\% \\ \hline
\hline
\end{tabular}
\end{center}
\end{table}

Using the branching fraction $Br(J/\psi\to\gamma\eta_c)=(1.3\pm0.4)\%$~\cite{pdg2002}, 
the $\eta_c$ branching fractions can be obtained. Table~\ref{etac-frac}
shows the BES results together with the PDG~\cite{pdg2002} and Belle~\cite{belle1,belle2} values.
The BES $Br(\eta_c\to \phi\phi)$ is smaller than
the current PDG value of $(7.1\pm 2.8) \times 10^{-3}$ and is consistent
with the Belle~\cite{belle2} and DM2~\cite{dm2} measurements within errors.
The branching fractions for $\eta_c\to K^\pm K^0_S \pi^\mp$ and $\eta_c\to p\bar{p}$ are
consistent with both the Belle\cite{belle1} and PDG values~\cite{pdg2002}.
The branching fractions for $\eta_c\to\pi^+\pi^-\pi^+\pi^-$ and $\eta_c\to K^+ K^-\pi^+\pi^-$
are consistent with the PDG values~\cite{pdg2002} within errors.

\begin{table}[htbp]
\caption{Branching fractions of the $\eta_c$ (the Belle results of $Br(\eta_c\to K^\pm K_{S}^{0}\pi^\mp)$
and $Br(\eta_c\to p\bar{p})$ are calculated from reference~\cite{belle2}).}
\label{etac-frac}
\begin{center}
\begin{tabular}{|l|c|c|c|}
\hline
Process                                & BES(\%)        & PDG02(\%) \cite{pdg2002}          & Belle(\%) \\ \hline
$Br(\eta_c\to K^+K^-\pi^+\pi^-)$       & $1.2\pm 0.4$   & $2.0^{+0.7}_{-0.6}$ & -         \\ \hline
$Br(\eta_c\to \pi^+\pi^-\pi^+\pi^-)$   & $1.0\pm 0.5$   & $1.2\pm0.4$         & -         \\ \hline
$Br(\eta_c\to K^\pm K_{S}^{0}\pi^\mp)$ & $1.7\pm 0.7$   & $\frac{1}{3}(5.5\pm1.7)$ & $\sim 1.8$ \\ \hline
$Br(\eta_c\to \phi\phi)$               & $0.25\pm 0.09$ & $0.71\pm0.28$       & $0.18^{+0.08}_{-0.06}\pm 0.07$  \\ \hline
$Br(\eta_c\to p\bar{p})$               & $0.15\pm 0.06$ & $0.12\pm0.04$       & $\sim 0.14$ \\ \hline
\hline
\end{tabular}
\end{center}
\end{table}

\newpage
\vspace{0.5cm}
   The BES collaboration thanks the staff of the BEPC and the IHEP computing center
for their efforts.
This work is supported in part by the National Natural Science Foundation
of China under contracts Nos. 19991480, 10225524, 10225525, the Chinese Academy 
of Sciences under contract No. KJ 95T-03, the 100 Talents Program of CAS 
under Contract Nos. U-11, U-24, U-25, and the Knowledge Innovation Project of 
CAS under Contract Nos. U-602, U-34 (IHEP); by the National Natural Science 
Foundation of China under Contract No.10175060 (USTC); and by the Department 
of Energy under Contract No. DE-FG03-94ER40833 (U Hawaii).

\end{document}